\documentclass[aip, sd, amsmath,amssymb, reprint]{revtex4-1}

\usepackage{graphicx}  
\usepackage{dcolumn}  
\usepackage{bm}  
\usepackage{dcolumn}  
\usepackage{amssymb}   
\usepackage{sidecap}
\usepackage[colorlinks=true, plainpages = false, citecolor = blue, urlcolor = blue, filecolor = blue]{hyperref}

\begin{document}

\preprint{AIP/123-QED}

\title[Applied Physics Letters]{Intrinsic Electron Mobility Limits in $\beta$-Ga$_2$O$_3$}

\author{Nan Ma}
\email{nanma@cornell.edu.}
\author{Nicholas Tanen}
\author{Amit Verma}
\affiliation{School of Electrical and Computer Engineering, Cornell University, Ithaca, NY 14853, USA.}

\author{Zhi Guo}
\author{Tengfei Luo}
\affiliation{Department of Aerospace and Mechanical Engineering, University of Notre Dame, Notre Dame, IN 46556, USA.}

\author{Huili (Grace) Xing}
\author{Debdeep Jena}
\email{djena@cornell.edu.}
\affiliation{School of Electrical and Computer Engineering, Cornell University, Ithaca, NY 14853, USA.}
\affiliation{Department of Materials Science and Engineering, Cornell University, Ithaca, NY 14853, USA.}

\date{\today}

\begin{abstract}
By systematically comparing experimental and theoretical transport properties, we identify the polar optical phonon scattering as the dominant mechanism limiting electron mobility in $\beta-$Ga$_2$O$_3$ to $<$200 cm$^2/$V$\cdot$s at 300 K for donor doping densities lower than $\sim$10$^{18}$ cm$^{-3}$.  In spite of similar electron effective mass of $\beta-$Ga$_2$O$_3$ to GaN, the electron mobility is $\sim$10$\times$ lower because of a massive Fr$\ddot{\text{o}}$hlich interaction, due to the low phonon energies stemming from the crystal structure and strong bond ionicity.  Based on the theoretical and experimental analysis, we provide an empirical expression for electron mobility in $\beta-$Ga$_2$O$_3$ that should help calibrate its potential in high performance device design and applications.  
\end{abstract}

\maketitle

$\beta-$Ga$_2$O$_3$ has recently emerged as an ultra wide-bandgap semiconductor $E_g = 4.6 - 4.9$ eV\cite{{PR65},{APL2010}} with 300 K electron mobility $\mu\sim$150 cm$^2/$V$\cdot$s,\cite{APEX15} attractive enough to potentially offer high-voltage electronic device performance\cite{SST2016} that is beyond the reach of the currently successful GaN and SiC platforms.  With the recent success in the synthesis of large-area bulk single crystal substrates and availability of nanomembranes,\cite{{SST2016},{JCG2013},{APL14}} $\beta-$Ga$_2$O$_3$ becomes a transparent conductive oxide (TCO) with significant potential.  It advances the field of oxide electronics from the traditional IGZO, perovskites (SrTiO$_3$, BaSnO$_3$, etc), and ZnO.\cite{{AM2012},{Nature2004},{APL2003}}  In this work, we have explored the physics of the intrinsic electron mobility limits in this material system and obtained expressions that should prove useful in device design. 
\begin{figure}[t]
\includegraphics[scale=0.25]{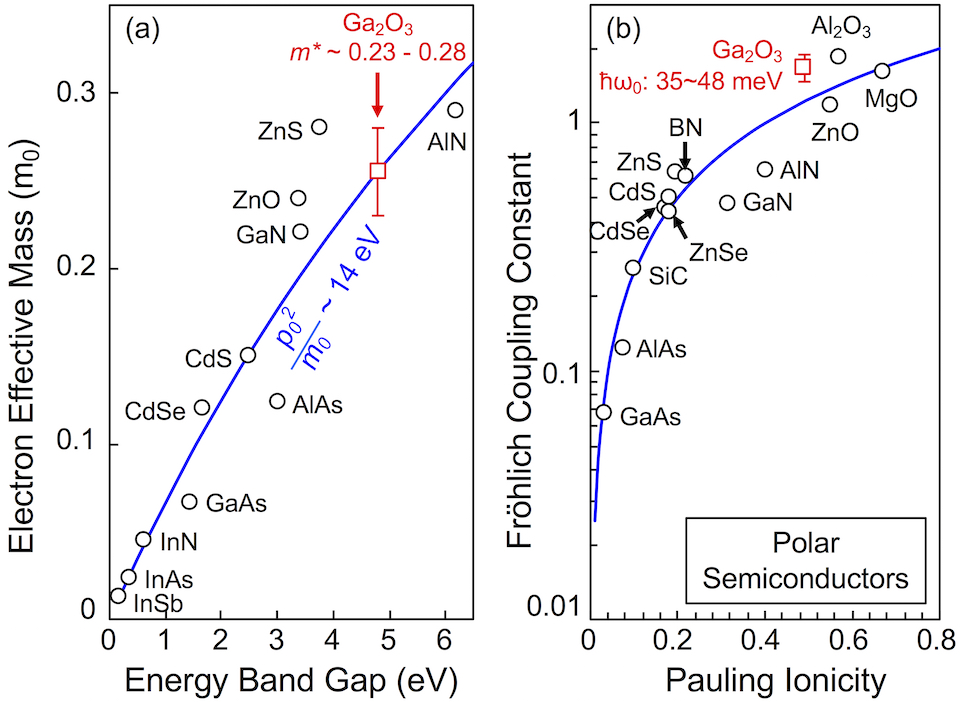}
\caption{\label{fig1} (a) Electron effective mass versus energy band gap and (b) Fr$\ddot{\text{o}}$hlich coupling constants versus Pauling ionicity for various compound semiconductors.}
\end{figure}

Since the Drude electron mobility $\mu=e\tau/m_c^{\star}$ is determined by the conduction band minimum (CBM) effective mass $m_c^{\star}$, electron charge $e$, and the low-field scattering rate $\tau$, we investigate $m_c^{\star}$ term first.  From standard ${\bf k} \cdot {\bf p}$ theory, $m_c^{\star}$ of sp$^3$-bonded direct gap semiconductors with the CBM at the $\Gamma$ point is related to $E_g$ by \cite{PRB77} $m_c^{\star}/m_0\sim(1+\frac{2p_0^2}{m_0}/E_g)^{-1}$, where $p_{0} \approx h / a_0 $ is the de-Broglie momentum of electrons at the Brillouin-zone edge $k = 2 \pi / a_0$ with $k$ the electron wavevector, $a_0$ the lattice constant, $h=2\pi \hbar$ the Planck's constant, and $m_0$ the free electron rest mass.  Figure \ref{fig1} (a) shows $m_c^*$ for various compound semiconductors as a function of $E_g$.\cite {{PRB01},{Adachi05}}  The solid blue line shows the ${\bf k} \cdot {\bf p}$ prediction with $2p_0^2/m_0\sim$14 eV.  Variations in lattice constant, Land$\acute{\text{e}}$ $g$ factor, slight indirectness of the bandgap, and the ionicity of the semiconductor can explain the slight deviations,\cite{PRB77} but the overall fit and the trend it predicts overrides these details -- with a large $E_g\sim$ 4.6 - 4.9 eV,\cite{{PR65},{APL2010}} $\beta-$Ga$_2$O$_3$ boasts a relatively low  $m_c^{\star} \sim 0.23-0.28m_0$,\cite{{SSC04},{APL2010}} as indicated in Fig. \ref{fig1} (a).  Unlike several other complex oxides and perovskites, where $|d\rangle $ or $| p\rangle $ orbital conduction bands lead to heavy $m_c^{\star}$, $\beta-$Ga$_2$O$_3$ has a small $m_c^{\star}$ because the CBM electron states derive from the hybridization of the Ga $|s\rangle$ orbitals.  Now because $m_c^{\star}$ of $\beta-$Ga$_2$O$_3$ is similar to that of GaN, one may initially expect the 300 K Drude mobility to be similar to bulk GaN ($\sim$1500 cm$^2/$V$\cdot$s).  However, the maximum experimentally measured 300 K electron mobility in bulk single-crystal $\beta-$Ga$_2$O$_3$ with little or no dislocations is $\sim$110-150 cm$^2$/Vs,\cite{{APL08},{CRT10},{JAP11},{APEX15},{APL14}} nearly {\em an order of magnitude} lower than GaN.  Potential electronic device applications of $\beta-$Ga$_2$O$_3$ beg the question whether the reported lower mobilities are intrinsic, or can be improved by eliminating extrinsic defects.  Answering this question is the subject of this work. 

We turn to the scattering rate to explain the difference in mobility between GaN and $\beta-$Ga$_2$O$_3$.  Because the Ga-O bond is strongly ionic,\cite{APL06} one can expect polar optical (PO) phonons to play an important role in limiting the room-temperature electron mobility, similar to GaN and GaAs.\cite{{PRB2000},{PRB86}}  The PO phonon energy $\hbar\omega_0$ intersects the electron bandstructure at the characteristic wavevector $k_0=\sqrt{2m_c^{\star}\omega_0/\hbar}$, which defines a characteristic Born-effective field\cite{Seeger} $E_0 = \frac{e^2}{4 \pi \varepsilon_0} \frac{k_0^2}{2} \left(\frac{1}{\varepsilon_{\infty}}-\frac{1}{\varepsilon_s}\right)$, where $\varepsilon_0$ is the vacuum permittivity, and $\varepsilon_s$ and $\varepsilon_{\infty}$ are the low- and high-frequency relative dielectric constant of the semiconductor.  For polar semiconductors, the strength of electron-PO phonon (e-PO) interaction is dictated by the dimensionless Fr$\ddot{\text{o}}$hlich coupling constant:\cite{AP54}   
 \begin{equation}
 \alpha_F=\frac{e E_0/k_0}{\hbar\omega_0}=\frac{e^2}{8\pi \varepsilon_0\hbar}\sqrt{\frac{2m_c^{\star}}{\hbar\omega_0}}\left(\frac{1}{\varepsilon_{\infty}}-\frac{1}{\varepsilon_s}\right)\label{eq1}
 \end{equation}
 
Figure. \ref{fig1} (b) shows $\alpha_F$ for several polar semiconductors plotted against their Pauling ionicity: $f_p=1-\text{exp}[-(X_{AB})^2/4]$\cite{Pauling}, where $X_{AB}$ is the difference of the electronegativity of the two elements in the bond.  The blue solid trend line shows the empirical relation $\alpha_F \sim 2.5\times f_p$, showing that $\alpha_F$ increases with $f_p$, i.e., strongly ionic bonds lead to stronger e-PO interaction.  Recent polarized reflectance measurements performed by Onuma et al.\cite{APL16} show that the lowest optical phonon modes $\hbar\omega_0$ in $\beta-$Ga$_2$O$_3$ are in the range of 35-48 meV, which leads to $\alpha_F\sim1.68\pm$0.21, as shown by the red open square in Fig. \ref{fig1} (b).  This indicates a massive Fr$\ddot{\text{o}}$hlich coupling, nearly 3$\times$ stronger than GaN.  Polaron effects, such as self-trapped holes due to their heavy band,\cite{PR69} can be expected to be strong in $\beta-$Ga$_2$O$_3$, renormalizing the electron effective mass $m_p^{\star}/m_c^{\star}\sim(1+\alpha_F/6)$,\cite{Seeger} which is used in the following transport properties study.

\begin{figure}[t]
\includegraphics[scale=0.25]{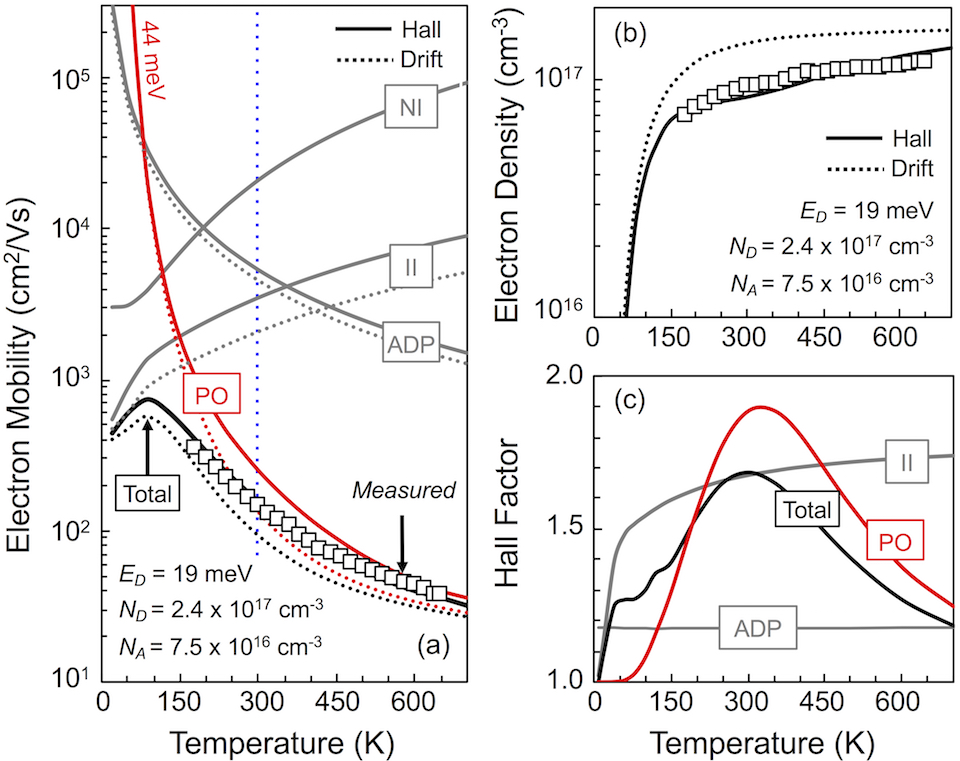}
\caption{\label{fig2} Temperature-dependent (a) electron mobility, (b) electron density, and (c) Hall factors in $\beta-$Ga$_2$O$_3$.  The solid and dashed lines in (a) and (b) indicate Hall and drift electron mobilities (densities), respectively.  The open squares indicate experimental results obtained using Hall-effect measurements.} 
\end{figure}

For non-degenerate dilute carrier densities at high temperatures ($T$) when $\hbar\omega_0/k_BT\ll1$, where $k_B$ is the Boltzmann constant, the PO phonon scattering limited electron mobility\cite{JAP1995} $\mu_{PO}\propto [\text{exp}(\hbar\omega_0/k_BT)-1]/\omega_0\alpha_F$.  From the 300 K $\mu_{PO}\sim$1500 cm$^2$/V$\cdot$s of bulk GaN, one can estimate $\mu_{PO}$ in $\beta$-Ga$_2$O$_3$ to be $\sim$88-141 cm$^2$/V$\cdot$s.  This value agrees with the reported $\mu\sim$110-150 cm$^2$/V$\cdot$s in $\beta$-Ga$_2$O$_3$ at 300 K.\cite{{APL08},{CRT10},{JAP11},{APEX15},{APL14}}  This simple analysis indicates that PO phonon is likely the dominant room-temperature scattering mechanism in $\beta$-Ga$_2$O$_3$ and rules out other possibilities of intrinsic scattering mechanisms. 

Hall-effect measurements were performed over a wide temperature range from 180 K to 650 K using a Lakeshore Hall system on unintentionally-doped Ga$_2$O$_3$ bulk substrates from Tamura Corporation.  These substrates were diced into 5 mm$\times$5 mm square pieces.  Ti/Pt Ohmic contacts were deposited on the four corners of the sample in Van de Pauw geometry followed by rapid thermal annealing (2 min, 480 $^{\circ}$C, Nitrogen atmosphere).  Figure \ref{fig2} (a) shows the measured temperature-dependent electron mobilities in $\beta-$Ga$_2$O$_3$, and Figure \ref{fig2} (b) shows the measured electron density, both as open squares.  The lines in Figure \ref{fig2} (a) show the calculated mobility resolved into the individual scattering mechanisms by ADP ($\mu_{ADP}$), ionized ($\mu_{II}$) and neutral impurity ($\mu_{NI}$) scattering, and the red line shows the polar-optical scattering limited mobility $\mu_{PO}$. These calculations are described after a brief discussion. The solid line in Figure \ref{fig2} (b) is the calculated Hall-effect carrier density $n_H$ based on the Hall mobility $\mu_H$ and the dashed line is $n$, corresponding to the drift mobility $\mu_d$.  

Figure \ref{fig2} (c) shows the calculated Hall-factor $r_H$ due to each scattering mechanism and the net $r_H$ with all scattering mechanisms considered, highlighting the difference between $\mu_H$ and $\mu_d$.  $N_D$, $N_A$, and the ionization energy $E_D$ are from fitting of the temperature-dependent $n$ using the neutrality condition: $n+N_A=N_D^+$.  For the current sample, $N_D=2.4\times10^{17}$ cm$^{-3}$, $N_A=7.5\times10^{16}$ cm$^{-3}$, and $E_D = 19$ meV.  Figure \ref{fig2} (a) indicates that extrinsic scattering from ionized and neutral impurity dominate the electron mobility at low temperatures up to $\sim$150 K.  Among intrinsic scattering mechanisms, $\mu_{ADP}$ is higher than the measured mobility by more than one order of magnitude over the entire temperature range.  Consequently the only mechanism that limits the electron mobility at high temperatures is PO phonon scattering.  The optical phonon energy $\hbar\omega_0$ that explains the measured temperature-dependent $\mu_H$ is $\sim$ 44 meV, which is very close to the lowest phonon energies from optical spectroscopy measurements reported by Onuma et al.\cite{APL16}  It lends credence to our claim of the dominance of PO phonon scattering in $\beta-$Ga$_2$O$_3$.

In a recent work, Parisini and Fornari suggested that intrinsic electron mobility in $\beta-$Ga$_2$O$_3$ is controlled by optical deformation potential (ODP) scattering.\cite{SST16}  It is known that ODP scattering plays an important role in non-polar crystals such as silicon and germanium, and disappears due to symmetry when the CBM is at the $\Gamma$ point and there are only two atoms in the primitive unit cell.\cite{{Ridley},{PR48},{PR56}}  $\beta-$Ga$_2$O$_3$, with CBM at the $\Gamma$ point, however has a monoclinic structure with ten atoms in the primitive unit cell, which leads to 30 phonon modes.\cite{SSC04}  Recent DFT calculation indicates that in $\beta$-Ga$_2$O$_3$, ODP scattering of electrons at low electric field are negligible compared to the PO phonon scattering,\cite{Buffalo16} similar to other polar semiconductors such as GaN and GaAs.\cite{{PRB2000},{PRB86}}  Thus, we do not consider ODP further in this work. 

To quantify the dominant role of e-PO scattering in $\beta$-Ga$_2$O$_3$, the lines of Fig \ref{fig2} were calculated using the relaxation-time approximation (RTA) solution of the Boltzmann transport equation (BTE).  Four scattering mechanisms are considered: ionized impurity (II), neutral impurity (NI), PO phonon, and acoustic deformation potential (ADP).  The material parameters used in the calculation are listed in Table \ref{table1}.  

\begin{table}
\caption{\label{table1} Material parameters of $\beta$-Ga$_2$O$_3$.}
\begin{ruledtabular}
\begin{tabular}{lcc}
 Parameter&Symbol&Value \\
\hline
Mass Density (g/cm$^3$) &$\rho$ &5.88\footnotemark[1]\\
Sound Velocity (cm/s) &$v_s$ &6.8$\times$10$^5$\footnotemark[2] \\
Acoustic Deformation Potential (eV)&$D_A$ & 6.9  \footnotemark[2] \\
Static Dielectric Constant& $\varepsilon_s$ & 10.2 \footnotemark[3] \\
High-frequency Dielectric Constant& $\varepsilon_{\infty}$ & 3.57\footnotemark[3] \\
CBM Electron Effective Mass (m$_0$)& $m_c^{\star}$ & 0.28\footnotemark [4] \\
PO Phonon Energy (meV)&$\hbar\omega_0$ & 44 (fitted) \\
\end{tabular}
\end{ruledtabular}
\footnotetext[1]{Ref.~\onlinecite{SST16}.}
\footnotetext[2]{Ref.~\onlinecite{{APL15},{ADP}}.}
\footnotetext[3]{Ref.~\onlinecite{{APL94},{JAP95},{APL02}}.}
\footnotetext[4]{Ref.~\onlinecite{APL2010}.}
\end{table}

The RTA solution of BTE gives the average electron drift mobility for carriers moving in 3-dimensions:\cite{Chihiro}
\begin{equation}
\mu_d=\frac{e\langle\tau_m\rangle}{m_p^{\star}}=\frac{e}{m_p^{\star}}\frac{2}{3k_BT}
\frac{\int dE\tau_mE^{3/2}f_0(1-f_0)}{\int dEf_0E^{1/2}},
\label{eq2}
\end{equation}
where $E$ is the electron energy, $f_0$ is the Fermi-Dirac distribution, and $\tau_m$ is the momentum relaxation time.  The Hall factor $r_H=\langle\tau_m^2\rangle/\langle\tau_m\rangle^2$ is calculated to obtain the Hall mobility $\mu_H=\mu_d\cdot r_H$ and the Hall electron density $n_H=n/r_H$, where $n$ is the mobile electron density.  $\tau_m$ due to each individual scattering mechanism is evaluated using Fermi's golden rule.  Phonon and neutral impurity scattering are assumed to be unscreened, while ionized impurity scattering is statically screened by free carriers with the reciprocal Debye screening length: $q_D=\sqrt{e^2n^{\star}/\varepsilon_0\varepsilon_sk_BT}$, where $n^{\star}$ is the effective screening carrier density:\cite{Wolfe}
$n^{\star}=n\cdot F_{-1/2}(\eta_C)/F_{1/2}(\eta_C)+(N_D-n-N_A)(n+N_A)/N_D$,
where $N_D$ and $N_A$ are the densities of donors and compensated acceptors.  $F_j(x)$ is the Fermi integral, $\eta_C=\frac{E_F-E_c}{k_BT}$, $E_F$ is the Fermi level, and $E_c$ the CBM energy.  $\tau_m$ of ionized impurity scattering is given by Brooks-Herring model:\cite{BrooksHerring} $\tau_{II}(k)^{-1}=\frac{e^4N_Im_p^{\star}}{8\pi\hbar^3\varepsilon_0^2\varepsilon_s^2k^3}[\text{ln}(1+b)-b/(1+b)]$, where $N_I=n+2N_A$, $b=4k^2/q_D^2$.  The scattering rate due to neutral impurities is given by:\cite{PR50} $\tau_{NI}^{-1}=80\pi N_n\hbar^3\varepsilon_0\varepsilon_s/e^2m_p^{\star}$, where $N_n=N_D-n-N_A$.  For ADP scattering: $\tau_{ADP}(k)^{-1}=D_A^2k_BTm_p^{\star}k/\pi\hbar^3\rho v_s^2$, parameters here are defined in Table \ref{table1}.  $\tau_m$ due to PO phonon scattering is
\begin{eqnarray}
&\tau_{PO\pm}(k)^{-1} &= \frac{e^2m_p^{\star}\omega_0}{16\pi\hbar^2 k^3\varepsilon_0}(\varepsilon_{\infty}^{-1}-\varepsilon_s^{-1})
(N_0+1/2\pm 1/2)\times \nonumber\\
&& \{4kq_0-2k_0^2[\text{ln}(k-q_0)+\text{ln}(k+q_0)]\},
\label{eq3}
\end{eqnarray}
where $q_0=\sqrt{k^2\mp k_0^2}$, and $N_0=[\text{exp}(\hbar\omega_0/k_BT)-1]^{-1}$ is the equilibrium phonon number. The $\lq\lq +"$ and $\lq\lq-"$ subscripts are for PO phonon emission and absorption.  The net momentum relaxation rate is obtained using Matthiessen's rule: 
$\tau_m^{-1}=\tau_{II}^{-1}+\tau_{NI}^{-1}+\tau_{ADP}^{-1}+\tau_{PO}^{-1}$.

\begin{figure}[t]
\includegraphics[scale=0.25]{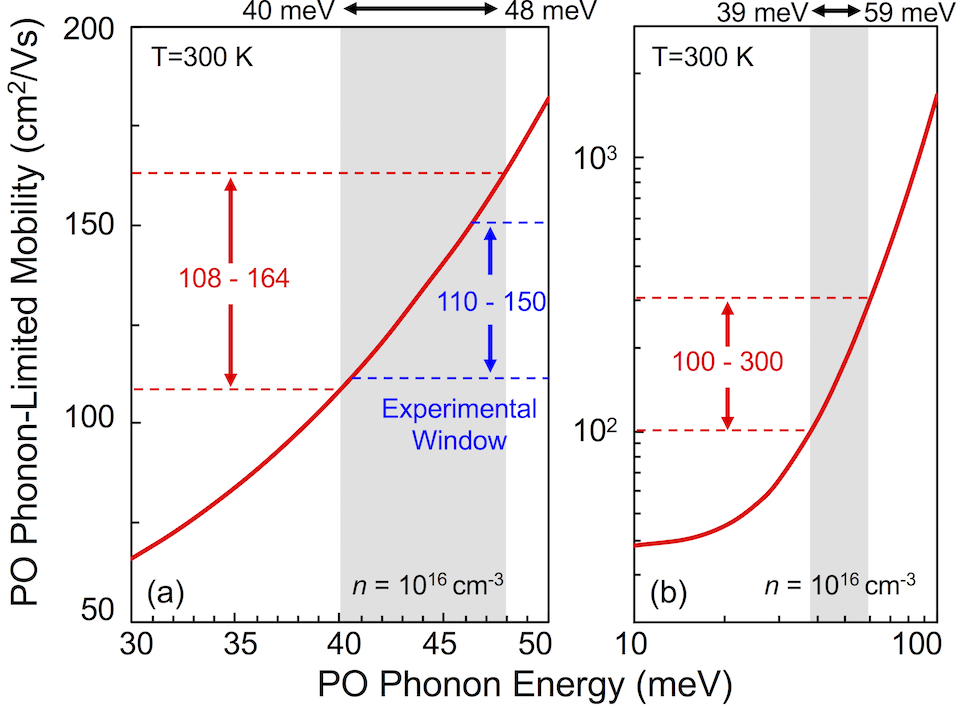}
\caption{\label{fig3}PO phonon limited electron mobility of $\beta-$Ga$_2$O$_3$ as a function of $\hbar\omega_0$ at room temperature.  Note that (a) is in linear scale and (b) is semi-log scale.}
\end{figure}
\begin{figure}[t]
\includegraphics[scale=0.25]{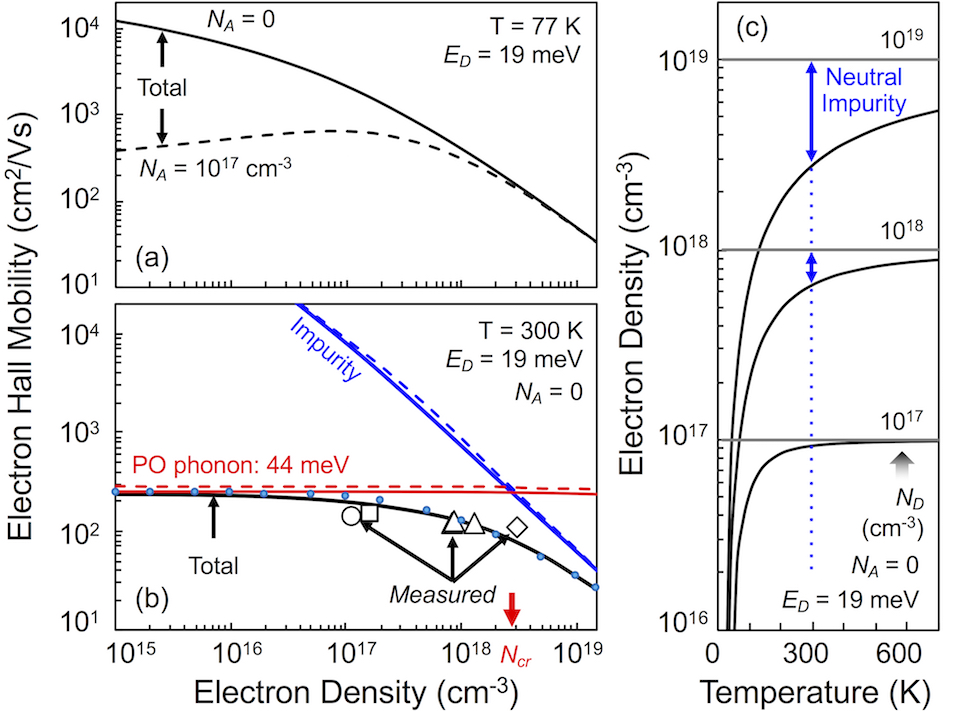}
\caption{\label{fig4}Electron mobilities as a function of donor concentration at (a) 77 K and (b) 300 K. The symbols show experimental results from different groups: $\circ$ Ref. \onlinecite{CRT10}, $\triangle$ Ref. \onlinecite{JAP11}, $\square$ Ref. \onlinecite{APEX15}, and $\diamond$ Ref. \onlinecite{APL14}. (c) Electron density as a function of temperature with different donor concentrations.}
\end{figure}
Figure \ref{fig3} shows the the sensitivity of $\mu_{PO}$ (drift) to the value of $\hbar\omega_0$ used in Eq. \ref{eq3} at 300 K.  When $\hbar\omega_0$ varies from 40 meV to 48 meV ($44\pm4$ meV),  $\mu_{PO}$ varies from 108 $\sim$ 164 cm$^2$/V$\cdot$s.  These values fully cover the reported experimental electron mobility values from various groups and are within acceptable experimental errors.  Even if we relax the $\mu_{PO}$ range to 100 $\sim$ 300 cm$^2$/V$\cdot$s (Fig. \ref{fig3} (b)), where the 300 cm$^2$/V$\cdot$s value is the qualitatively estimated intrinsic mobility predicted by Sasaki et al.,\cite{APEX12} the corresponding $\hbar\omega_0$ ranges from $\sim$ 39 to 59 meV.  Therefore $\hbar\omega_0$ in $\beta$-Ga$_2$O$_3$ that limits electron transport at room temperature is inferred to be not far from $\sim$ 44 meV as we have extracted from our experimental results.  
With the e-PO interaction established as the dominant intrinsic scattering mechanism, it is important to study the relative importance of impurity scattering on electron mobility.  Figure \ref{fig4} (a) and (b) show the calculated $\mu_H$ as a function of $N_D-N_A$ at 77 K and 300 K.  For 77 K, the solid and dashed lines show mobilities with $N_A=0$ and $N_A=10^{17}$ cm$^{-3}$.    As seen in Fig. \ref{fig4} (a), intrinsic electron mobility higher than $\sim$ 10,000 cm$^2$/V$\cdot$s can be achieved in very clean samples with $N_D \lesssim$10$^{15}$ cm$^{-3}$.  However, introducing $N_A$ severely decreases electron mobilities through ionized impurity scattering.  When $N_A$=$10^{17}$ cm$^{-3}$ and $N_D-N_A=10^{15}$ cm$^{-3}$, the mobility is reduced to $\sim$ 350 cm$^2$/V$\cdot$s.  This explains the experimentally observed low-temperature low electron mobilities,\cite{{JAP11},{APEX15}} and also much higher mobilities\cite{Kumagai} ($\sim$ 7,000 cm$^2$/V$\cdot$s) in cleaner samples.  In Fig. \ref{fig4} (b), the black line shows the net $\mu_H$ with all four scattering mechanisms considered, the red solid line indicates $\mu_{PO}$ with $\hbar\omega_0$ = 44 meV and the blue solid line show mobility limited by ionized and neutral impurity scattering.  The open symbols are experimental results from various groups.  Figure \ref{fig4} (c) shows $n$ as a function of temperature for various $N_D$.  The arrows denote the density of neutral impurities.  The heavier the doping, the more unionized impurities are introduced.  Therefore both ionized and neutral impurity scattering become increasingly important in determining electron mobilities as temperature decreases and as the doping concentration increases.  

To estimate $\mu_H$ around or higher than 300 K, we derived the following expressions:
\begin{widetext}
\begin{equation}
\mu_{PO}=\frac{e\sqrt{m_p^{\star}}}{\pi^2\hbar^3\omega_0n\alpha_Fk_BT}\times\left[\frac{5}{3N_0}\int_0^{\hbar\omega_0} dEE^{3/2}f_0(1-f_0)+\frac{7}{6(N_0+1)}\int_{\hbar\omega_0}^{\infty} dEE^{3/2}f_0(1-f_0)\right]r_{po},\label{eq4}
\end{equation}
\end{widetext}
\begin{equation}
\mu_{II}=\frac{128\sqrt{2\pi}\varepsilon_s^2\varepsilon_0^2(k_BT)^{3/2}}{e^3\sqrt{m_p^{\star}}N_I[\text{ln}(1+\beta^2)-\beta^2/(1+\beta^2)]}r_{ii},\label{eq5}
\end{equation} 
\begin{equation}
\mu_{NI}=e^3m_p^{\star}/80\pi\hbar^3\varepsilon_s\varepsilon_0N_n,\label{eq6}
\end{equation}
where $\beta=2\sqrt{6m_p^{\star}k_BT}/q_D\hbar$, $r_{po}$ and $r_{ii}$ are the Hall factors of PO phonon scattering and ionized impurity scattering, respectively.  At 300 K, the empirical expressions $r_{po}=1.876\{1+[n(\text{cm}^{-3})/2\times10^{19}]^{1.22}\}$ and $r_{ii}=0.99/1+[n(\text{cm}^{-3})/3.2\times10^{18}]^{0.68}+0.8$ are used to describe the doping dependence of Hall factors.  The first and second terms in the brackets of Eq. \ref{eq4} correspond to the PO phonon absorption and emission, respectively. 

The dashed lines in Fig. \ref{fig4} (b) show the net $\mu_H$ calculated using Eq. \ref{eq4}-\ref{eq6}.  At 300 K, PO phonon scattering dominates the mobility for $N_D$ lower than a critical doping density $N_{cr}$ determined by the crossover condition $\mu_{PO}\approx(\mu_{II}^{-1}+\mu_{NI}^{-1})^{-1}$.  When $N_D>N_{cr}$, neutral impurity introduced by heavy doping severely degrades the electron mobility.  For 300 K, $\mu_{PO} \sim$ 250 cm$^2$/V$\cdot$s and $N_{cr} \sim 2.76\times10^{18}$ cm$^{-3}$.  For power devices that typically operate at high temperatures, $\mu_H$ is captured by the empirical expression:
\begin{equation}
\mu_H = \frac{ 56\left\{\text{exp}[\frac{508}{T(\text{K})}]-1\right\} }{\left\{1+ \frac{ N_D(\text{cm}^{-3})}{[T(\text{K})-278]2.8 \times 10^{16}} \right\}^{0.68}} \frac{\text{cm}^2}{\text{V}\cdot\text{s}}.
\label{eq7}
\end{equation} 
The 300 K electron mobility calculated using Eq. \ref{eq7} is shown by blue dots in Fig. \ref{fig4} (b).  This expression, which is valid for 300 K$\lesssim T\lesssim$500 K, offers a useful guideline for experiments and is easily embedded in device modeling.  

In summary, we studied the intrinsic electron mobility limits in \textit{n}-doped $\beta-$Ga$_2$O$_3$.  We find an extremely strong Fr$\ddot{\text{o}}$hlich interaction in the material, fueled by the high ionicity of the chemical bonds, and the low optical phonon energies.  A PO phonon energy of $\sim$44 meV was deduced from the transport properties.  The measured room-temperature electron mobility is dominated by PO phonon scattering for low doping densities, and thus has approached the intrinsic mobility limits.  Though it is difficult to change the PO phonon energy, one could investigate strain as a potential tool.\cite{APL2015}   Moreover, similar to III-V semiconductors, the formation of a two-dimensional electron gas (2DEG) at (AlGa)$_2$O$_3/$Ga$_2$O$_3$ heterojunctions is possible with modulation doping.  The electron mobility in such a 2DEG is expected to be higher than the bulk due to the relaxation of the momentum conservation in the direction perpendicular to the interface during the electron-PO phonon scattering process, the elimination of neutral impurity potential, and the powerful exponential reduction of ionized impurity scattering by remote doping.  

\begin{acknowledgments}
This work was supported by NSF DMREF program (Award Number 1534303).  The authors thank Guru Khalsa for useful discussions. 
\end{acknowledgments}


\begin{thebibliography}{99}
\bibitem{ADP}
The deformation potential $D_\alpha^i$ of a particular band $i$ induced by a certain phonon mode $\alpha$ through electron-phonon interaction is given by $D_\alpha^i=\Delta V_i/(\Delta l_\alpha/l_0)$, where $\Delta V_i$ is the energy change of band $i$ relative to that in the equilibrium structure, $l_0$ is the lattice constant of the equilibrium structure, and $\Delta l_\alpha$ is the lattice constant change due to a structural perturbation from phonon mode $\alpha$. The calculation is carried out by first displacing the atomic coordinates in the optimized cell from their equilibrium positions by a small amount through following the eigenvector of a certain phonon mode. The band-structure calculation is then performed on a series of such deformed structures with different $\Delta l_\alpha$ ($-0.01l_0$ to $0.01l_0$). The deformation potential for the band gap $\Delta V_{bg}=\Delta V_{CBM}-\Delta V_{VBM}$ is thus obtained by a linear fit of  $\Delta V_{bg}$ versus $\Delta l_\alpha/l_0$. The first-principles calculation details were described in \onlinecite{APL15}
\end{thebibliography}
\end{document}